\documentclass[prl,11pt,tightenlines,letterpaper]{revtex4}
\usepackage{graphicx}

\begin{document}

\begin{center}
\textbf{Kinetic exchange vs. room temperature ferromagnetism in
diluted magnetic semiconductors}\end{center}
\ \\
J. Blinowski$^1$, P. Kacman$^2$ and T. Dietl$^2$\\
$^1$Institute of Theoretical Physics, Warsaw
University, ul.~Ho\.za 69, PL-00-681 Warszawa, Poland\\
$^2$Institute of Physics, Polish Academy of Sciences,
al.~Lotnik\'ow 32/46, PL-02-668 Warszawa, Poland
\begin{flushleft}
\textbf{Abstract}

\hspace*{0.3in} Guided by the internal-reference rule and the
known band offsets in III-V and II-VI diluted magnetic
semiconductors, we discuss the feasibility of obtaining p-type
conductivity, required for the carrier-induced ferromagnetism, as
well as the cases for which the doping by shallow impurities may
lead to the ferromagnetism driven by the double exchange. We
consider the dependence of kinetic exchange on the p-d
hybridization, on the electronic configurations of the magnetic
ions, and on the energies of the charge transfer between the
valence band of host materials and the magnetic ions. In the case
of Mn-based II-VI compounds, the doping by acceptors is necessary
for the hole-induced ferromagnetism. The latter is, however,
possible without any doping for some of Mn-, Fe- or Co-based III-V
magnetic semiconductors. In nitrides with Fe or Co carrier-induced
ferromagnetism with $T_{\mbox{\small{C}}}>300$~K is expected in
the presence of acceptor doping.

\section{}
\hspace*{0.3in}The discovery of the ferromagnetic order in
Mn-based III-V \cite{Ohno92,Ohno96,Esch97,Mats98} and II-VI
diluted magnetic semiconductors (DMS) \cite{Haur97,Ferr01} started
intensive studies of these materials and their layered structures
\cite{Ohno99a,Diet01a}. The recently demonstrated phenomena, such
as spin-injection from (Ga,Mn)As contacts \cite{Ohno99b}, tuning
of magnetic properties by an electric field \cite{Ohno00}, and
large tunneling magnetoresistance \cite{Tana01}, brought closer
the idea of new semiconductor spin devices combining complementary
features of semiconductor and magnetic systems. For this purpose
it is, however, of great importance to increase the Curie
temperatures of ferromagnetic DMS. Much technological effort has
been recently put in the search for different DMS exhibiting
ferromagnetism at room temperature. The hope for finding such
materials was awaken when the Curie temperatures above room
temperature was predicted for materials containing light anions
\cite{Diet00}. To explain the origin of the ferromagnetism in
tetrahedrally coordinated DMS as well as the values of the Curie
temperatures the mean-field Zener model was employed
\cite{Diet00}. In this model, the ordering of spins results from
the p-d kinetic exchange interaction between the magnetic ions and
the delocalized or weakly localized holes. Accordingly, the large
value of the exchange energy $N_0 \beta$ is one of the crucial
conditions for high-temperature ferromagnetism, along with the
large (of the order of few $10^{20}$~cm$^{-3}$) hole and Mn
concentrations. We note that in n-type DMS, only very low Curie
temperatures can be expected because of small s-d exchange energy
and low density of states in the conduction band \cite{Diet97}.
Indeed, no ferromagnetism was detected above 1~K in (Zn,Mn)O:Al
\cite{Andr01}.

\hspace*{0.3in}The values of $N_0 \beta$ are well known for many of the II-VI DMS
which have been extensively studied in the last two decades
\cite{Kacm01,Koss93}. In contrast, the available information
on this constant in III-V DMS concerns primarily GaAs with
Mn$^{2+}$ ions -- for this material several values, ranging from
$-0.9$~eV \cite{Bhat00} and  $-1$~eV \cite{Szcz01} to $-4.5$ eV
\cite{Sanv01} have been suggested. At present, the most reliable
seems to be the value $N_0 \beta=-1.2$ eV inferred from the
photoemission data \cite{Okab98}. The latter value (scaled by the
inverse volume of the elementary cell) was used  for predicting the
Curie temperatures for various III-V DMS with Mn$^{2+}$ (i.e.,
d$^5$-configuration) ions \cite{Diet00}.

\hspace*{0.3in}In this paper, we  discuss the position of electronic states
introduced by transition-metal impurities in II-VI and III-V
compounds from the view point of ferromagnetism mediated either by
band carriers or double exchange. We show also a strong dependence
of the p-d exchange energy on the number and configuration of the
electrons residing on the d shell of various magnetic ions.

\hspace*{0.3in}According to the internal reference rule \cite{Caldas,Lang88}, the
positions of states derived from d shell of magnetic ions do not
vary across the entire family of the II-VI or III-V compounds if
the valence band offsets between different compounds are taken into
account. In Fig.~1 (left panel) we present the collected data for the II-VI DMS
containing various transition metal ions. We have denoted by D(0/+)
and A(0/$-$) the donor and acceptor levels (i.e., the lower and upper
Hubbard bands), which correspond to the transformation of the doubly
ionized magnetic ions M$^{2+}$ into M$^{3+}$ and into M$^{1+}$ ions,
in their ground states, respectively. By D$^*(0/+)$ and A$^*(0/-)$ we
denote the excited donor and acceptor levels, with lower spin than
the corresponding ground states. Energies of  these levels enter the
formulae that determine the exchange energy (see Eqs.~1 and 2).

\begin{figure}
\hspace*{-1cm}\includegraphics[scale=0.62]{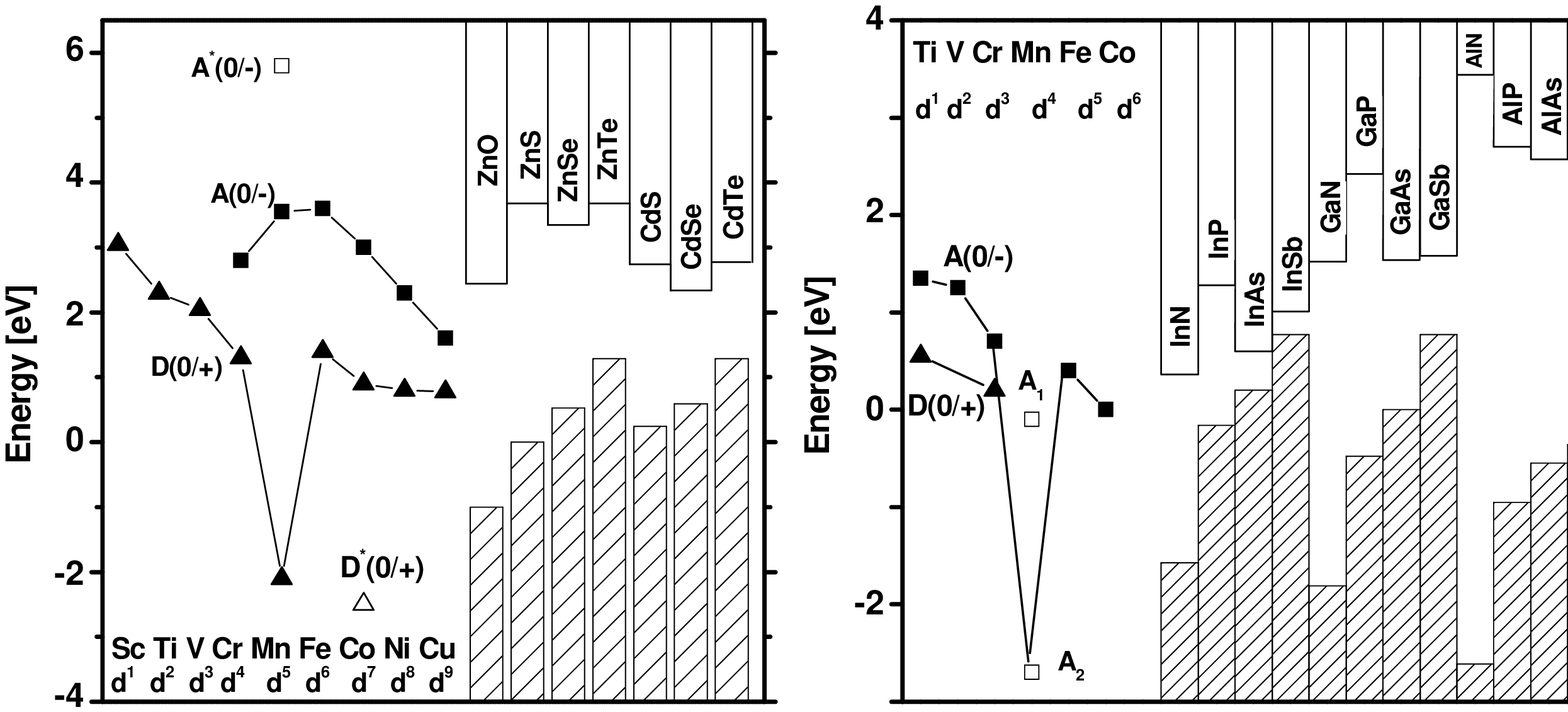}
\caption{Approximate positions of transition metals levels
relative to the conduction and valence band edges of II-VI (left
panel) (after
\cite{Koss93,Lang88,Vogl85,Zung86,Soko87,Heim90,Sze, Wei98}) and
III-V (right panel) (after \cite{Lang88,Zung86,Vurg01}) compounds. By
triangles the d$^{N}$/d$^{N-1}$ donor and by squares the
d$^N$/d$^{N+1}$ acceptor states are denoted; in (left) the solid
symbols represent the ground states, whereas the open symbols the
excited final states with lower spin values. In (right) the states
labeled A$_1$ and $A_2$ correspond to the d$^4$/d$^5$ acceptors as
given by spin-resonance studies in GaP:Mn \cite{Krei96} and by
photoemission in (Ga,Mn)As \cite{Okab98}, respectively.
\label{refrule}}
\end{figure}

\hspace*{0.3in}Some information about the possibility of the hole induced
ferromagnetism in II-VI DMS can be directly obtained from Fig.~1.
First of all, we observe that the doping with shallow acceptors is
necessary to obtain the p-type materials as the acceptor levels of
all the transition metal ions lay well above the top of the valence
band of every II-VI compound. Moreover, one can see that for all
transition metals with $N \neq 5$ (perhaps apart from Co in the
tellurides) also the donor level lays above the valence band and,
thus, the appearance of the band holes (even with an additional
acceptor doping) is excluded.  From this point of view, the
Mn$^{2+}$ ($N=5$) ion is quite unique, since its donor level is
situated well below the top of the valence band of all the II-VIs,
giving the chance for p-type DMS. Indeed, in (Cd,Mn)Te and
(Zn,Mn)Te doped with N or P the hole-induced ferromagnetism was
observed between 1 and 3~K \cite{Haur97,Ferr01,Andr01}. In view of
the above discussion, we note that the recently reported
room-temperature ferromagnetic behavior of n-type (Zn,Co)O
\cite{Ueda01} and (Zn,V)O \cite{Saek01} layers cannot be attributed
to the carrier induced mechanism described by the Zener or RKKY
model. The data may suggest that double exchange involving
hopping of electrons over acceptor levels (upper Hubbard band) is
involved. We note, that for V ions in II-VIs also  the
superexchange was predicted to be ferromagnetic, \cite{Blin96}.

\hspace*{0.3in}It should be, however, noted at this point that the internal
reference rule may serve only for the illustration of chemical
trends and not for extracting the precise values of the ionization
energies. Moreover, in DMSs other states related to the magnetic impurities 
may also appear, e.g.\cite{Schnei87}. Strong p-d hybridization can lead to
binding of a hole in a Zhang-Rice polaron, which then gives rise to
an additional state in the band gap \cite{Beno92,Diet01b}.
Furthermore, if the d$^N$/d$^{N-1}$ donor level resides above the
bottom of the conduction band, the ground state corresponds to a
hydrogenic-like donor level d$^{N-1}$+e located below the band
edge, as observed in CdSe:Sc \cite{Glod94}. Similarly, if the
acceptor state is located under the top of the valence band, the
ground state corresponds to a hydrogenic-like acceptor d$^{N+1}$+h,
not to the d$^N$ state. Obviously, energies of hydrogenic-like
states follow the band edges, and by no means are described by the
internal reference rule. This appears to be the case of the Mn
related levels in III-V compounds \cite{Diet01b}. These effects
cause some ambiguity concerning the nature of states observed
experimentally in magnetically doped semiconductors.

\hspace*{0.3in}To examine chemical trends in III-V DMS we present in Fig.~1 (right panel) the
literature data concerning valence band offsets and energy levels
of transition metal impurities for the III-V semiconductors. Here
D(0/+) and A(0/-) denote again the donor and acceptor states,
which, in contrast to the situation in II-VI DMS, correspond to the
transformation of the triply ionized magnetic ions M$^{3+}$ into
M$^{4+}$ and into M$^{2+}$ ions, respectively. The  energies of all
but Mn transition metal impurity levels (solid symbols in Fig.~1)
were taken from Ref.~\cite{Lang88}. The position of Mn acceptor
A(0/$-$) level with respect to the valence band edges of the
III-V-s is not conclusively established.  According to optical and
EPR experiments in GaP:Mn, the Mn d$^4$/d$^5$ level is located at
0.4~eV above the valence band edge (A$_1$ in Fig.~1)
\cite{Krei96}. On the other hand, the analysis of photoemission in
(Ga,Mn)As \cite{Okab98} suggests this level to reside deep ($\sim
2.7$~eV) below the valence band edge (the A$_2$ position). The latter
position of the Mn d$^4$/d$^5$ level is consistent also with the
occupation number of the Mn d-states close to 5, inferred from
the x-ray magnetic circular dichroism \cite{Ohla00}, and with the
results of the LSDA+U computations \cite{Park00}.  In the following
we show that these two (A$_1$ and A$_2$) positions, when used as
the input values for the internal reference rule, lead to quite
different conclusions about the ferromagnetism in Mn based III-V
DMS.

\hspace*{0.3in}From Fig.~1 one can see that the chances of obtaining the p-type in
III-V are better than in II-VI DMS. With the position A$_2$ of the
Mn d$^4$/d$^5$ acceptor level the large concentration of holes in
the valence band can be obtained in all presented III-V compounds
without any additional doping.  Thus, the conditions required for
the ferromagnetism in the Zener model are fulfilled in all these
compounds \cite{Diet01b}, exept perhaps AlN. In GaSb and
InSb the same can be expected even if the Mn d$^4$/d$^5$ level
position is A$_1$ and also for Co-based DMS. The antimonides are,
however, less promising for high temperature ferromagnetism,
because of their relatively large lattice constants and,
consequently, weak p-d hybridization. For Fe in GaSb and InSb and
for Co and Mn  (A$_1$) in GaAs and InAs the internal reference rule
implies the acceptor level slightly below the valence band edge,
i.e., the p-type with relatively low concentrations of delocalized
holes can be expected. In such case the Fermi level, after reaching
the acceptor level position, would remain pinned to it. The
coexistence of the M$^{2+}$ and M$^{3+}$ ions would open an
additional channel of the exchange interaction, namely, the
ferromagnetic double exchange, as proposed \cite{Akai98} for
uncompensated (In,Mn)As. In materials in which the acceptor
A(0/$-$) level is situated within the band gap, the p-type can be
obtained only by shallow acceptor doping. This requires, however,
the donor level to reside below the valence band edge. According to
Fig.~1, in the case of Ti and Cr ions, the p-type samples seem to
be excluded.

\hspace*{0.3in}Within the second order perturbation theory the kinetic exchange
can be described in terms of hybridization-induced virtual transitions
of electrons between the band and the ionic d-shell. In the case of II-VI
diluted magnetic compounds it was shown that the kinetic exchange
mechanism depends crucially on the electronic configuration of the
magnetic ions \cite{Blin92} and the charge transfer energies for
the transitions from the band onto the magnetic ion and vice-versa
\cite{Blin96}. For the p-band in the vicinity of the $\Gamma$
point, the transitions only to/from the ionic  t$_{2g}$ orbitals
are allowed. For tetrahedrally coordinated DMS with various magnetic ions,
the contribution to the kinetic exchange from the transitions
involving singly occupied orbitals is proportional to the exchange
constant \cite{Kacm01}:
\begin{equation}
N_0\beta_N=-\frac
{{V_{pd}}^2}{S}\left[\frac{1}{E_{N-1}^{S-1/2}+\epsilon_p -
E_{N}^{S}}+\frac{1}{E_{N+1}^{S-1/2}-\epsilon_p - E_{N}^{S}}\right]
\end{equation}
whereas the contribution involving empty $t_{2g}$ orbitals is
proportional to:
\begin{equation}
N_0\gamma_N=\frac
{{V_{pd}}^2}{(S+1/2)}\left[\frac{1}{E_{N+1}^{S+1/2}-\epsilon_p -
E_{N}^{S}}+\frac{1}{E_{N+1}^{S-1/2}-\epsilon_p - E_{N}^{S}}\right]
\end{equation}
where $V_{pd}$ can be related to the Harrison inter-atomic matrix
elements: $V_{pd}=\frac{4}{3}(V_{pd\sigma}-2V_{pd\pi}/\sqrt{3})$ ;
$E_{N}^{S}$ is the energy of the ion with $N$ d-electrons and the
total spin $S$; $\epsilon_p$ is the energy of the electron at the
top of the p-band. It has been also shown that for ions with
all t$_{2g}$ orbitals occupied by the same number of electrons, the
kinetic exchange leads to the Kondo-like Hamiltonian, with the
exchange constant $N_0\beta$ for ions with d$^5$, d$^6$ or  d$^7$
electronic configurations and with $N_0\gamma$ for the d$^1$ and
d$^2$ ions \cite{Blin92}. For other electronic configurations the
kinetic exchange Hamiltonian is more complex and depends on the
ionic orbital degrees of freedom -- still, for the d$^3$ and d$^4$
ions the Kondo-like form of the Hamiltonian can be restored by the
static Jahn-Teller effect (provided the cubic symmetry is in
average conserved) with the exchange constant being an appropriate
combination of $N_0\beta$ and $N_0\gamma$.

\hspace*{0.3in}The values and even the signs of all exchange constants depend on
the charge transfer energies in the denominators in Eqs.~(1) and
(2), which are related to the distances between the donor or
acceptor levels and the edge of the valence band.

\hspace*{0.3in}As predicted previously \cite{Diet00}, among the II-VI DMS the best
candidate for the hole-induced, high temperature ferromagnetic
system seems to be (Zn,Mn)O. In this compound one expects the value
of the exchange energy to be considerably greater than in the other
II-VI DMS. This is due to the small elementary cell of ZnO, which
results in the large hybridization constant $V_{pd}$. The
phenomenological scaling used in \cite{Diet00} leads to the value
of $N_0\beta$ in ZnO:Mn  equal to $-2.48$~eV. A comparable value,
$N_0\beta=-2.1$~eV, was recently obtained in Ref.~\cite{Namb01}
with the parameters entering to Eq.~(1) determined by the best fit
to the photoemission spectra. The value obtained with the charge
transfer energies given by the internal reference rule in the
Fig.~1 and the Harrison hybridization parameters is even larger,
i.e., $-3.2$~eV. 

\hspace*{0.3in}In a given host material the values of the
$N_0\beta$ constant are known to be considerably larger for Co than
for Mn ions. This is primarily due to the smaller value of spin $S$
in the denominator of Eq.~(1), since the charge transfer energies
are similar in both cases (note that for Co$^{2+}$ ($N=7$) the
charge transfer energy  not from the D(0/+) but from D$^*(0/+)$
enters Eq.~(1)). Unfortunately, this enhancement of the exchange
constant is compensated by the decrease of magnetic susceptibility
(proportional to $S^2$). Thus, one should expect similar Curie
temperatures for the p-type tellurides with Mn and Co ions -- relatively
low, due to their large lattice constants.

\hspace*{0.3in}In III-V DMS the strongest hybridization favoring the high temperature ferromagnetism
is expected in phosphides and nitrides. In these compounds the
acceptor levels of Fe and Co are situated above the valence band.
Thus, in p-type samples the charge state of these ions would remain
the same as of the host cations and their electronic configurations
are d$^5$ and d$^6$, respectively. The estimated in \cite{Blin01}
values of $N_0\beta$ [$-2.8$~eV in (Ga,Fe)N and $-3.1$~eV in
(Ga,Co)N] suggest that in these compounds, the ferromagnetism with
$T_{\mbox{\small{C}}}>300$~K could be expected, provided  that a
high concentration of holes can be introduced. Preliminary data on
(Ga,Fe)N, with the low Fe content (of the order
$10^{19}$~cm$^{-3}$) have already revealed ferromagnetic properties
below 100~K \cite{Akin00}.

\hspace*{0.3in}To evaluate the prospects of obtaining ferromagnetic Mn-based
nitrides and phosphides with high Curie temperature the
discrimination between the positions A$_1$ and A$_2$ of the
d$^4$/d$^5$ acceptor level is of primary importance. This is
because, with the gap position A$_1$ an additional doping would be
necessary to obtain the p-type material, and the electronic
configuration of the ions would be d$^4$, which is {\em not}
favorable for high $T_{\mbox{\small{C}}}$. In this configuration,
due to a partial cancellation of two terms, the antiferromagnetic
$N_0\beta$ and the ferromagnetic $N_0\gamma$, the kinetic exchange
would be considerably reduced \cite{Blin01}. The first samples of
(Ga,Mn)N reported in the literature \cite{Zaja01} seemed to be of
n-type with ionized (d$^5$) Mn acceptors. The position of the
acceptor level at about 1~eV above the valence band edge, deduced
from optical measurements \cite{Wolo01}, suggests the above
described scenario in the hypothetical p-type samples. This
position does not fulfill, however, the internal reference rule
with either $A_{1}$ or A$_{2}$ positions. It may be rather ascribed
to the binding energy of holes in Zhang-Rice polaron states bound
to d$^5$ Mn acceptors \cite{Diet01b}. This interpretation selects
the $A_2$ position of the Mn d$^4$/d$^5$ acceptor level, deep in
the GaN valence band. With this position and, therefore, the d$^5$
configuration of Mn ions in the p-type material, the Curie
temperatures well exceeding 300~K were predicted within the Zener
model \cite{Diet00,Diet01b}, in consistence with the recently
reported indications of ferromagnetic transitions in (Ga,Mn)N
\cite{Theo01,Reed,Sono01}.
\section{}
\textbf{Acknowledgments:}

\hspace*{0.3in}Valuable discussions with H. Ohno and F.
Matsukura as well as partial support of the FENIKS project (EC:
G5RD-CT-2001-00535) and by Foundation for Polish Science are
gratefully acknowledged.
\section{}

\end{flushleft}
\end{document}